\documentclass[a4paper,12pt]{article}
\usepackage{amsmath,amsfonts,amssymb,cite}

\begin{document}

\begin{center}
{\Large\bf On holographic relation between radial meson trajectories and deconfinement temperature\footnote{Presented at HADRON 2019.}}
\end{center}

\begin{center}
{\large S. S. Afonin}
\end{center}

\begin{center}
{\it Saint Petersburg State University, 7/9 Universitetskaya nab.,
St.Petersburg, 199034, Russia\\
E-mail: s.afonin@spbu.ru}
\end{center}

\begin{center}
{\large A. D. Katanaeva}
\end{center}

\begin{center}
{\it University of Barcelona, Mart\'{i} i Franqu\`{e}s 1,\\
Barcelona, 08028,  Catalonia, Spain\\
E-mail: katanaeva@fqa.ub.edu}
\end{center}

\begin{abstract}
The interrelation between the deconfinement temperature of hadron medium and parameters of radial Regge trajectories within the bottom-up holographic models for QCD is scrutinized. We show that the lattice data on the deconfinement temperature can yield a powerful restriction on the spectrum of excited mesons and glueballs within the framework of holographic approach. The best phenomenological agreement and theoretical self-consistency are achieved if the scalar meson $f_0(1500)$ is considered as the lightest glueball.
\end{abstract}



\section*{}

One of primary questions in the QCD phase diagram is to calculate the critical
temperature $T_c$ at which hadronic matter undergos
a transition to a deconfined phase.
Within the bottom-up holographic approach this type of studies was initiated by Herzog in Ref.~\cite{Herzog:2006} and continued by
many authors (see, {\it e.g.}, a brief review in Ref.~\cite{estim_deconf}).
In this approach, the gravity part of a $5D$ model is assumed to come from a dual description of
gluodynamics and can be used to study thermodynamic properties of original $4D$ gauge theory.
The deconfinement is related to the Hawking--Page phase transition
between a low temperature thermal Anti-de Sitter (AdS) space and a high temperature
black hole in the AdS/QCD models.

The estimation of $T_c$, in general, is model dependent.
Traditionally one fixes the hadron parameters from the vector meson sector due to a relatively rich experimental data on light vector mesons.
However, there seems to be no theoretical reason why the vector meson spectra should
be preferred.

In the present study, we will argue that the scalar glueball (and its radial excitations) is much better option for
fixation of the model parameters.
Our main arguments can be shortly formulated as follows.
(i) Phase diagram can be studied in pure gluodynamics.
 Since the holographic approach is defined in the large-$N_c$ limit of gauge theories
 where the glueballs dominate over the usual mesons and baryons
 the gluodynamics must dictate the overall mass scale and thereby the major contribution
 to the deconfinement temperature $T_c$.
(ii) Using isospectrality concept~\cite{Vega_Cabrera} we show that the predicted values of $T_c$ are more stable
for scalar glueballs than for vector mesons.
(iii) Phenomenological lattice reasons: Numerical values of $T_c$ determined from the scalar glueballs on the lattices fit much better the lattice results for  $T_c$.

These argument are scrutinized in our paper~\cite{estim_deconf} (see also~\cite{add6,add7}).


Let us introduce the $5D$ holographic action with a universal gravitational part and some matter part,
\begin{eqnarray}
S &=& \int d^4xdz\sqrt{-g} f^2(z)\left(\mathcal L_{gravity} + \mathcal L_{matter} \right),\\
\mathcal L_{gravity} &=& -\frac1{2k_g} \left(\mathcal{R}-2\Lambda\right).
\end{eqnarray}
Here $g_{MN}$ ($g=\det g_{MN}$) represents an AdS related metric, $k_g$ is a factor proportional to the $5D$ Newton constant, $\mathcal{R}$ is the Ricci scalar and $\Lambda$ is the cosmological constant.
The choice of the dilaton background $f(z)$ dictates a particular holographic model.
They differ as well by the interval the $z$ coordinate spans. We assume $z\in [0,z_{max}]$,
though $z_{max}=\infty$ is possible and will be of the main interest in the present work (the soft-wall (SW) background).

The holographic calculation of critical temperature is related to the leading contribution in the large-$N_c$ counting,
that is the $\mathcal L_{gravity}$ part scaling as $\frac1{2k_g} \sim N_c^2$ while $\mathcal L_{matter}$ scales
as $N_c$.
According to Ref.~\cite{Herzog:2006} the deconfinement in AdS/QCD occurs as the Hawking--Page phase transition that is a transition between different gravitational backgrounds.
We call the order parameter of this transition $\Delta V$.

$V$'s are the free action densities evaluated on different backgrounds corresponding to two phases.
The confined phase is given by the thermal AdS of radius $R$ and defined by the general AdS line element
\begin{equation}
ds_{Th}^2=\frac{R^2}{z^2}\left(dt^2-d\vec{x}^2-dz^2\right),
\end{equation}
with the time direction is restricted to the interval $[0, \beta]$.
The metric of the Schwarzschild black hole in AdS describes the deconfined phase and is given by
\begin{equation}
ds_{BH}^2=\frac{R^2}{z^2}\left(h(z)dt^2-d\vec{x}^2-\frac{dz^2}{h(z)}\right),
\end{equation}
where $h(z)=1-(z/z_h)^4$ and $z_h$ denotes the horizon of the black hole.
The corresponding Hawking temperature is related to the horizon as $T=1/(\pi z_h)$.

The cosmological constant in $5D$ AdS is $\Lambda=-6/R^2$ and
both these metrics are the solutions of the Einstein equations. They provide the same value of the Ricci scalar $\mathcal R = -20/R^2$.
Hence, the free action densities differ only in the integration limits,
\begin{align}
V_{\text{Th}}(\epsilon)&=\frac{4R^3}{k_g} \int_0^\beta dt\int_\epsilon^{z_{max}}dzf^2(z) z^{-5},\\
V_{\text{BH}}(\epsilon)&= \frac{4R^3}{k_g}  \int_0^{\pi z_h}dt\int_\epsilon^{\min(z_{max},z_h)}dzf^2(z)z^{-5}.
\end{align}
The two geometries are compared at $z=\epsilon$ where the periodicity in the
time direction is locally the same, {\it i.e.} $\beta=\pi
z_h\sqrt{h(\epsilon)}$. Then, we may construct the order parameter for the phase
transition,
\begin{equation}
\label{deltaV}
\Delta V = \lim_{\epsilon\rightarrow0}\left(V_{\text{BH}}(\epsilon)-V_{\text{Th}}(\epsilon)\right).
\end{equation}
The thermal AdS is stable when $\Delta V>0$, otherwise the
black hole is stable. The condition $\Delta V=0$ defines the critical temperature $T_c$ at which the
transition between the two phases happens.
Eqn.~(\ref{deltaV}) yields $z_h$ as a function of the model dependent parameters -- $z_{max}$ and/or those possibly
introduced in $f(z)$. We must invoke the matter sector $\mathcal L_{matter}$ to give physical meaning to
these parameters and to connect $T_c$ to a particular type of a holographic model.

As was recently noticed in Ref.~\cite{Vega_Cabrera}, the soft-wall background is not fixed by the form of linear spectrum as
one can find an infinite number of one-dimensional potentials leading to identical spectrum of normalized modes.
The corresponding family of potentials is referred to as isospectral potentials.

In brief, the problem for mass spectrum in the bottom-up holographic models can be reduced to a one-dimensional Schr\"{o}dinger equation
\begin{equation}
\label{Schrod}
-\psi''_n(z)+\widehat{\mathcal V}(z) \psi_n(z) =M^2(n)\psi_n(z),
\end{equation}
where $\widehat{\mathcal V} (z)$ is the Schr\"{o}dinger potential which depends on the 5D dilaton background
$f^2(z)$, metric, and spin.
A particular form of the Schr\"{o}dinger potential defines the eigenvalues of Eqn.~(\ref{Schrod}) and hence
the mass spectrum $M(n)$. In the case of SW models it is a potential similar to the one that appears
when considering the radial part of the wavefunction of a $2D$ harmonic oscillator system.

According to Ref.~\cite{Vega_Cabrera} and references therein, there exists
the following isospectral transformation between $\mathcal V_J(z)$ and  $\widehat{\mathcal V}_J(z)$,
\begin{equation}
\widehat{\mathcal V}_J(z) = \mathcal V_J(z) -2\frac{d^2}{dz^2}\ln[I_J(z)+\lambda].
\end{equation}
This technique allows one to generate a family of dilaton functions $f(z)$ appearing in $\widehat{\mathcal V}_J(z)$,
each member assigned to the value of the parameter $\lambda$. The case of $\lambda=\infty$ corresponds to the original $\mathcal V_J(z)$.
The function $I_J(z)$ is defined through the ground eigenstate of $\mathcal V_J$, $\psi_0$, and is given by
\begin{equation}
I_J(z)\equiv\int\limits_0^z\psi_0^2(z')dz'.
\end{equation}
Different $\lambda$ provide slightly different form of the potential, but the eigenvalues of Eqn.~(\ref{Schrod})
and, hence, the spectrum remain the same.

The main problem we studied can be formulated as follows: Does isospectrality entail isothermality (i.e. identical predictions for $T_c$)?
Generically the answer is negative. But we found one important exception: If model parameters are fixed from the
scalar glueball channel within the generalized SW holographic model of Ref.~\cite{GSW} (which is able to reproduce accurately the glueball
radial spectrum) then an isospectral family of models leads to almost identical predictions for the deconfinement temperature. The
typical predictions lie in the range $T_c\simeq 175 \pm 15$~MeV which agrees very well with modern unquenched lattice estimations.
The further details are contained in Refs.~\cite{AK2018, AK2019}.

\end{document}